\documentclass[twocolumn,prl,superscriptaddress,showpacs]{revtex4}
\usepackage{epsfig}
\usepackage{times}
\usepackage{epsf}
\usepackage{graphicx}
\date{\today}

\begin{document}
\title{Binding of holons and spinons in the one-dimensional anisotropic
  $t$-$J$ model} 
\author{Jurij \v{S}makov}
\author{A. L. Chernyshev}
\author{Steven R. White}
\affiliation{Department of Physics and Astronomy, University of
  California, Irvine, California 92697, USA}
\begin{abstract}
We study the binding of a holon and a spinon in the one-dimensional 
anisotropic $t$-$J$ model using
a Bethe-Salpeter equation approach, exact diagonalization, and density matrix
renormalization group methods on chains of up to 128 sites. We
find that holon-spinon binding changes dramatically
as a function of anisotropy parameter $\alpha=J_\perp/J_z$:  
it evolves from an exactly deducible impurity-like
result in the Ising limit to an exponentially
shallow  bound state near the isotropic case. A 
remarkable agreement between the theory and numerical results 
suggests that such a change  is controlled 
by the corresponding evolution of the spinon energy spectrum. 
\end{abstract}
\pacs{71.10.Fd, 71.10.Li, 75.10.Pq, 75.40.Mg} 

\maketitle
In the one-dimensional $t$-$J$ and Hubbard models, spin and charge dynamics are
independent, leading to the well-known effect of spin-charge separation:
the splitting of the electron (hole) into spinon and holon
excitations that carry only spin and only charge, respectively \cite{LiebWu}. 
Considerably less is known about interactions among these 
%``elementary'' 
excitations. Recently, it was shown that in the supersymmetric 
$t$-$J$ model, spinons and holons attract each other \cite{Bernevig},
but
%, contrary to a naive expectation,  
this does not result in a 
bound state, indicating that the spinon-holon interaction is non-trivial. 
%----------------------------------------------------------------------
\begin{figure}[b]
\includegraphics*[width=0.8\hsize,scale=1.0]{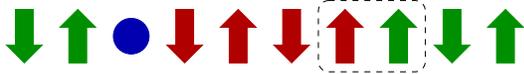}
\caption{(Color online). 
A hole  in the Ising AF (circle) moved by four sites from origin.
 The spinon is indicated by the dashed box.
\label{fig:spinon_holon}
}
\end{figure}
%----------------------------------------------------------------------

On the other hand, there is a limit in which spinon and holon
do form a bound state: the $t$-$J_z$ model \cite{sorella-1d}. 
Here, the half-filled system is an
Ising antiferromagnet (AF). Removing a spin and
moving the hole away from origin 
increases the net magnetic energy by $J_z/2$ due to 
creation of a domain wall in the AF order -- an
immobile spinon (Fig. \ref{fig:spinon_holon}).  Once moved, the
hole also carries an AF domain wall and is a free holon
with dispersion $\epsilon^0_k\!=\!-2t\cos k$.
Since recombination with the spinon lowers the energy
 of the system, the holon can be considered as moving 
freely except at the origin where it is subject to an 
effective attractive potential $V\!=\!-J_z/2$. 
Clearly, such a potential in 1D always leads to
a bound state. The single-hole Green's function 
can be found exactly: 
$G(\omega)\!=\!\left[J_z/2\pm\sqrt{\omega^2-4t^2}\right]^{-1}$.
Its lowest pole gives the binding energy of the 
spinon-holon bound state:
\begin{equation}
\label{ising-delta}
\Delta = 
2t\left(1-\sqrt{1+J^2_z/16t^2}\right),
\end{equation}
in which holon is confined in the vicinity of the spinon with the
localization length $\ell\!\propto\! \Delta^{-1/2}$.
This leads to the question: how does the Ising limit, in which the 
spin and charge are bound, evolve into the isotropic 
limit where they separate? 
This question, together with the expectation that
the study of anisotropic model 
can shed light on the general problem of spinon-holon 
interactions, motivates this work. 

In this Letter we address this problem by a
detailed theoretical and numerical study of the anisotropic
 $t$-$J$ model using the Bethe-Salpeter equation (BSE), exact
diagonalization (ED), and density matrix renormalization group (DMRG) \cite{dmrg}
methods on systems of up to 128 sites. 
Although the binding becomes too weak to be detected numerically near the 
isotropic limit, we believe that 
a holon and a spinon form a bound state for any value of
anisotropy $\alpha=J_\perp/J_z$ less than one. 
Given a remarkable agreement between the theory and numerical results,
we argue that the binding is largely controlled by the
spinon energy spectrum. Qualitatively, at
$\alpha\ll 1$ the holon is confined in
the vicinity of the impurity-like spinon, while already for
$\alpha\agt 0.5$ the spinon is quasi-relativistic and is only 
weakly bound to the holon.
Such a change in the pairing character results in a dramatic
decrease of the binding energy at larger $\alpha$.
In addition,
using a small-$\alpha$ analysis, we provide insight into 
the kinematic structure of the spinon-holon interaction that 
offers a simple explanation for having an attraction but no
bound state in the isotropic limit of the model.

Using standard notation, we define a
generalization of the 1D $t$-$J$ model,
${\mathcal H}=\sum_{i,j}{\mathcal H}_{i,j}$, with 
\begin{eqnarray}
\label{htot}
{\mathcal H}_{ij}=-t\sum_{\sigma}c^{\dagger}_{i\sigma}
c^{\phantom{\dagger}}_{j\sigma}
+\frac{J_z}{2}\left(S^z_i S^z_j+\alpha S^{+}_iS^{-}_{j}
-\frac{n_in_j}{4}\right),
\end{eqnarray}
where $i$ and $j=i\pm1$ are the
neighboring sites.
This model interpolates between the $t$-$J_z$ model 
($\alpha=0$) and the isotropic $t$-$J$ model ($\alpha=1$). 
We set $t=1$.
%Starting from the $t$-$J_z$ limit and 
%turning on transverse spin-spin interactions 
%results in a qualitative change: due to
%spin flips the spinon is no longer stationary. 
%Intuitively, the bound state will survive
%at small $\alpha$ as the spinon will 
%still be ``heavy'' compared to the holon. 
%An investigation of how the spinon-holon binding evolves
%as a function of $\alpha$, is the main topic of this Letter.

\begin{figure}[t]
\includegraphics*[width=1.0\hsize,scale=1.0]{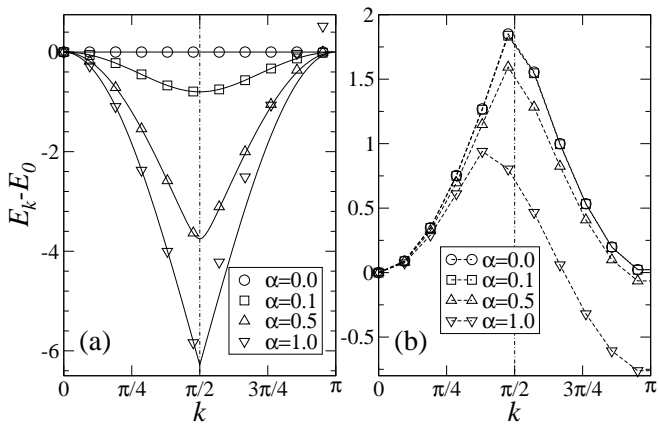}
\caption{Lowest energies $E_k\!-\!E_0$ 
 vs $k$ in an $L\!=\!21$ chain with (a) zero and (b) one hole, $J_z\!=\!4.0$. 
Solid lines in (a) is the spinon dispersion from BA 
\cite{johnson-excitations-xyz}; in (b) the lines are guides to the eye.
\label{fig:dispersion}
}
\end{figure}

\begin{figure}[t]
\includegraphics*[width=0.9\hsize,scale=1.0]{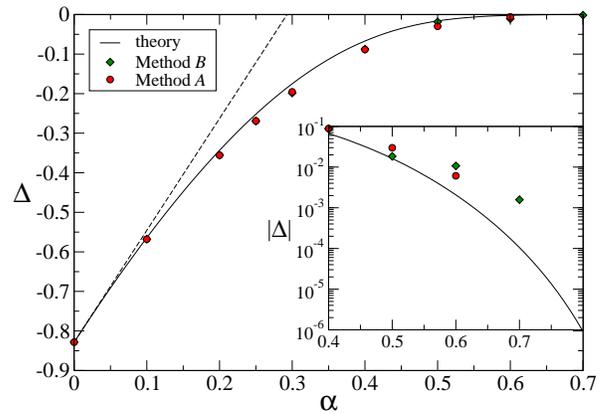}
\caption{(Color online). Theoretical (lines) 
and numerical results 
by method $A$ (circles) and method $B$ (diamonds) for $\Delta$ vs $\alpha$, 
$J_z=4.0$. Inset: $|\Delta|$ for $\alpha \ge 0.4$ on a semi-log plot.
\label{fig:binding}
}
\end{figure}

To obtain the binding energy numerically, we have performed
calculations of the ground state energies (GSEs)
using both ED and DMRG techniques. 
While only small system sizes are
accessible by ED (up to $L\!=\!23$ sites in our case), it allows us to
calculate the lowest energy in each momentum sector
(Fig. \ref{fig:dispersion}) and is also important in analyzing 
various finite-size effects. 
Using DMRG \cite{dmrg} we have calculated GSEs of
systems of up to $L\!=\!128$ sites using periodic boundary conditions (PBCs), which greatly
increases the numerical effort required. Up to $m\!=\!1400$ states per block were kept 
in the finite system method, with corrections applied to the density matrix to accelerate
convergence with PBCs \cite{singlesite}.
 The binding energy $\Delta$ for the
infinite system was obtained using two different extrapolation
methods described below. 
In the range of $\alpha$ where the calculation of $\Delta$ is
reliable ($\alpha\!\lesssim\!0.6$), we found excellent agreement between
the numerical data from both methods, and theoretical results, based
on BSE.
A summary of the results for $J_z\!=\!4.0$ is presented in
Fig. \ref{fig:binding}; results for other values of $J_z$
are qualitatively similar.
%----------------------------------------------------------------------
%----------------------------------------------------------------------

\emph{Numerics.}\ For a given $J_z$ and $\alpha$, 
the binding energy is
\begin{equation}
\label{binding0}
\Delta = E_{p}-E_{s}-E_{h},
\end{equation}
where $E_x$, $x\!=\!p,s,h$, is the GSE of the $L\!=\!\infty$ chain 
with the spinon-holon pair, the spinon, and the holon, all relative 
to GSE of the $XXZ$ chain, respectively. 
It is well known \cite{shiba-ogata-review,zotos-tbc}
that the energies of individual spinon and holon can be 
obtained from GSEs of odd-$L$ periodic chains 
with no holes and one hole, respectively. The even-$L$ periodic 
chain with one hole contains the spinon-holon pair. 
A consistent 
finite-size definition of $\Delta_L$ for even $L$ is:
\begin{equation}
\label{bindingb}
\Delta_L=E^0_L+E^1_L-\frac12\left(E^0_{L-1}+
E^0_{L+1}+E^1_{L-1}+E^1_{L+1}\right),
\end{equation}
where $0$ and $1$ refer to the number of holes. 
It can be used directly to obtain a sequence of binding
energies $\Delta_L$ for different $L$, which can then be extrapolated
to $L\!=\!\infty$ limit. We will call this approach ``method
$B$''. An alternative and substantially more precise way to estimate
$\Delta_\infty$ is: (i) to obtain $E_x(L)$ by subtracting 
the extensive part of energy $\tilde\varepsilon_\alpha L$ from 
corresponding GSEs ($\tilde\varepsilon_\alpha$ is
the energy per site of an infinite $XXZ$ chain, known from 
Bethe-Ansatz (BA) \cite{yang-xxz}), (ii) extrapolate $E_x(L)$ 
to the infinite system size, and 
(iii) use Eq. (\ref{binding0}). We will refer to this
approach as ``method $A$''. 
In this method  we use the exact $L=\infty$ value 
for the spinon energy $E_{s}$
that is also known from BA \cite{johnson-excitations-xyz}, thus 
completely eliminating one of the sources of the 
finite-size effects. 

\begin{figure}[t]
\includegraphics*[width=1.0\hsize,scale=1.0]{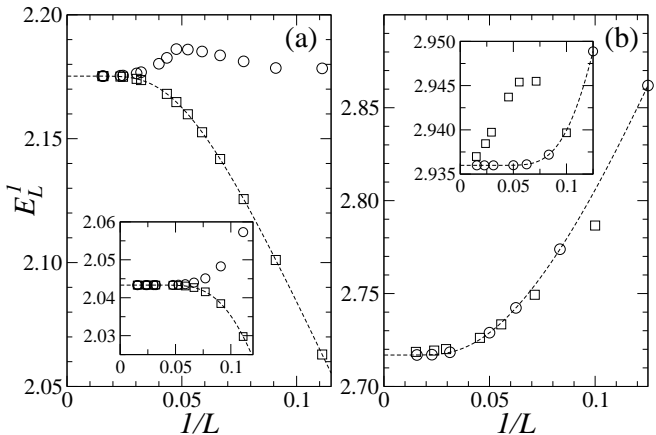}
\caption{The $4$-even (circles) and $4$-odd (squares) branches for
(a) holon  and (b) pair energies vs $1/L$. $J_z=4.0$
and $\alpha=0.4$ ($\alpha=0.25$, insets).  Dashed
curves are the fits described in text.
\label{fig:extrap}
}
\end{figure}

\emph{Method A}.\ One of the problems is the ``staggered'' behavior of the 
holon and pair energies vs $L$, 
meaning that the data must be separated into two branches, defined
by whether $L$ for the pair ($L-1$ for the holon) is divisible 
by 4. We refer to them as ``$4$-even'' and ``$4$-odd'' branches, respectively  
(see also Ref. \onlinecite{Bernevig}).
Within each branch energy depends on the size smoothly.  
We found for both holons and pairs that although
one of the branches may exhibit a non-uniform dependence on system size, 
the other one is always well-behaved, see Fig. \ref{fig:extrap}.
Thus, even though we have fewer points in the ``good'' branch, 
the quality of extrapolation using it is vastly improved.

In fact, for the holon data, we can further advance this success by
doubling the 
number of points in each branch 
using the ``twisted'' boundary conditions (BCs)
 \cite{zotos-tbc}. 
The GS of a holon in an infinite chain is degenerate for
$k=0$ and $k=\pi$. For a finite $L$, this
degeneracy is lifted: in $4$-even chains $E_{0}$ 
is shifted up and $E_{\pi}$ is higher in the $4$-odd ones 
(see Fig. \ref{fig:dispersion}).  
The staggering occurs because the $k$-space of odd-$L$ chains 
does not contain the $k=\pi$ point.  
This can be mitigated by 
flipping the sign of the hopping integral $t$, which 
shifts the momentum by $\pi$. 
Thus, we can construct the GSEs for both branches 
in every odd-$L$ chain by combining the data as follows:
\begin{equation}
\left.
\begin{array}{l}
\text{mod}(L-1, 4) = 0, t=+1\\
\text{mod}(L-1, 4) \not = 0, t=-1
\end{array}
\right\}\ \ 4\text{-even branch}.
\end{equation}
The complementary data goes into $4$-odd branch. An example of such
reconstruction, along with the extrapolation to the infinite size, is
shown in Fig. \ref{fig:extrap}(a).
%----------------------------------------------------------------------
%----------------------------------------------------------------------

For the spinon-holon pair data, the separation into two branches also 
allows for a smooth $E$ vs $1/L$ scaling. Further improvement
using twisted BCs
can be achieved by multiplying $t$ with an {\it a priori} unknown
$L$-dependent complex phase factor \cite{zotos-tbc}. 
While such BCs can be easily handled by ED, 
our DMRG code requires extensive modifications to support them. 
Results featuring phase-adjusted data will be 
presented elsewhere \cite{elsewhere}.
%----------------------------------------------------------------------
\begin{table}[b]
\begin{tabular}{|c|c|c|c|}
\hline
Branch & $\text{mod}(L,4)$ & $\text{sign}(t)$ for $E^1_{L-1}$ &
$\text{sign}(t)$ for $E^1_{L+1}$\\ 
\hline
% le-he 
$B1$ & 0 & -1 & +1 \\
% le-ho 
$B2$ & 0 & +1 & -1 \\
% lo-he 
$B3$ & 2 & -1 & +1 \\
% lo-ho 
$B4$ & 2 & +1 & -1 \\
\hline
\end{tabular}
\caption{
Subdivision of the numerical data for $\Delta(L)$ into different branches
due to finite size effects in method $B$.
\label{tab:branches}
}
\end{table}
%----------------------------------------------------------------------

Having significantly improved the scaling quality of the available GSE data,
we need to choose the functional form for the $E$ vs $1/L$ fit. 
We find that at not too large $\alpha$ GSEs approach $E_\infty$
exponentially fast. 
Notably, similar exponential convergence of the 
energy of the finite $XXZ$ chain at large $L$ was derived by
Woynarovich and de Vega (WdV) using BA 
\cite{vega-woynar-size-corr}. 
Thus, we attempt to fit both holon and pair data with 
$E_L\!=\!E_\infty\!+\!e^{-\lambda L}P_n(1/L)$,
where $P_n(x)$ is a polynomial of order $n$. 
Remarkably, the values 
of $\lambda$ we found for holons are in close agreement 
with WdV ones, so we used them in our holon fits.
For pairs there is no such correlation,
so we retain $\lambda$'s as free parameters. 
Example of pair energy extrapolation is shown in Fig. \ref{fig:extrap}(b).
%----------------------------------------------------------------------
%----------------------------------------------------------------------

Finally, we are able to obtain high-precision values 
for the $L\!=\!\infty$ binding energies in the range 
$0\!\le\!\alpha\!\le\!0.5$ using (\ref{binding0}). 
The data for $J_z=4.0$ are shown in Fig. \ref{fig:binding}. 
By studying the quality of the
fits and the variation of $\Delta_\infty$ depending on the fit
type, we estimate the error to be less than 3\% for $\alpha\!=\!0.5$ 
and negligible for smaller values of $\alpha$.
For $\alpha\!>\!0.5$ parameter 
$\lambda$ drops below $1/L_{max}$, so the corresponding fits become 
less reliable. For $\alpha\!=\!0.6$ 
the error is of the order of 10\%, and for 
$\alpha=0.7$ it exceeds $100$\%.
In principle, our results at larger $\alpha$
can be improved by increasing $L$ while keeping the DMRG precision the same 
($\sim\!10^{-4}t$). 
However, as one can see from the inset in Fig. \ref{fig:binding}, 
the binding energy itself diminishes rapidly and it is likely to hit the 
DMRG accuracy around $\alpha\sim 0.8$. 

%----------------------------------------------------------------------
\begin{figure}[t]
\includegraphics*[width=0.95\hsize,scale=1.0]{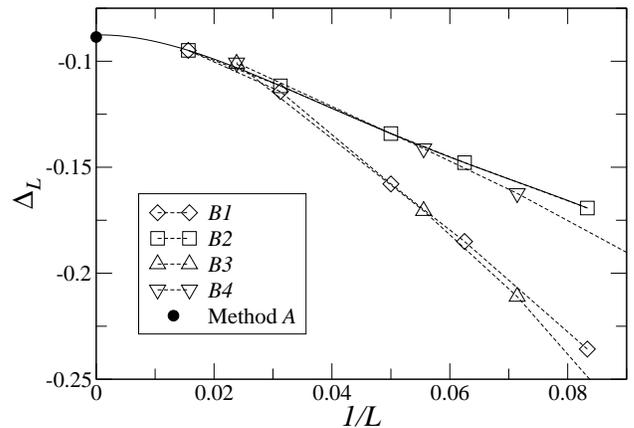}
\caption{$\Delta_L$ data vs $1/L$ in
method $B$, $J_z=4.0$ and $\alpha=0.4$. Branches are defined in 
Table \ref{tab:branches}. Dashed lines are guides to the eye,
solid line is the $4$-th order polynomial fit of the $B2$ branch.
\label{fig:branches}
}
\end{figure}
%----------------------------------------------------------------------

\emph{Method B}.\ The ``direct'' extrapolation of $\Delta_L$
using Eq. (\ref{bindingb}) (method $B$), does not rely on
prior knowledge of BA results. Thus, it 
provides an important validity test for the results of 
method $A$. The $\Delta_L$ data is, again, staggered, and 
has to be separated into the $4$-even and $4$-odd branches. We also have 
a freedom in choosing the sign of $t$ 
for the holon energies $E^1_{L-1}$ and $E^1_{L+1}$. As a
result, data splits into four different branches, defined in Table
\ref{tab:branches}.
An example of size dependence for different branches is shown in
Fig. \ref{fig:branches}. 
The best choice, corresponding to the ``good'' holon branch in method $A$,
is branch $B2$. The results of extrapolation of $B2$ data using polynomial fit 
of maximum order are shown in Fig. \ref{fig:binding}. They are in close 
agreement with method $A$. 
Inspection of the terms in (\ref{bindingb}) reveals that most significant
finite-size effects in method $B$ come from the spinon energies $E^0_{L\pm 1}$ 
(eliminated in method $A$). 
Even though method $B$ serves as
an important validity test, its relative error is
always larger than that of method $A$.

\emph{Theory.}  
Generally, the binding energy in 1D should scale 
as $-V^2m$, where $V$ is the interaction
and $m$ is a particle mass.  For the Ising limit 
this estimate gives $\Delta\!\sim\!-J_z^2/t$ 
in agreement with the exact answer (\ref{ising-delta}).
%At $\alpha\!=\!0$ the spinon is impurity-like and the 
%spinon-holon binding is easily solved \cite{elsewhere} to give
%(\ref{ising-delta}). 
To extend our approach to the finite $\alpha$,  we use BSE formalism 
\cite{LL} in which two particles with dispersions 
$\epsilon_k$ and $\omega_q$, and interaction  $V_{q,q'}$
create a bound state if their scattering amplitude has a pole.
At the pole, BSE can be simplified 
to the Schr\"odinger equation for
the pair wavefunction $\chi(q)$:
\begin{equation}
\label{BSE}
\chi(q) = \frac{1}{E-\epsilon_{q+P}-\omega_q}\sum_{q'} 
V_{q,q'}\chi(q'),
\end{equation}
where $P$ is the total momentum of the pair, and the pair energy
$E\!=\!\Delta\!+\!\epsilon_0\!+\!\omega_0$ 
is the binding energy $\Delta$ relative to the 
band minima of the the particles
$\epsilon_0\!=\!min[\epsilon_k]$ and $\omega_0\!=\!min[\omega_q]$.
In the Ising limit, $\epsilon_k\!=\!-2t\cos k$, 
$\omega_q\!=\!J_z/2$, and 
$V_{q,q'}\!=\!-\omega_0$, so Eq. (\ref{BSE}) yields 
a dispersionless ($P$-independent) bound state with
$\Delta$ given by (\ref{ising-delta}). 

At $\alpha\!>\!0$ spinon becomes mobile and the interaction 
$V$  and holon dispersion change.
The holon mass renormalization was found to be insignificant  
throughout the anisotropic regime $0\!<\!\alpha\!\leq\!1$ \cite{zotos-tbc}.
On the other hand, spinon dispersion changes 
drastically as it evolves from the gapped, immobile excitation 
with energy $\omega_q\!=\!J_z/2$ at $\alpha\!=\!0$  to relativistic, 
gapless excitation with $\omega_q\!=\!J_z(\pi/2)|\cos q|$ 
in the isotropic limit. 
The spinon dispersion for $XXZ$ model, shown by solid lines in Fig. 
\ref{fig:dispersion}(a), is known exactly from BA: 
$\omega_q\!=\!c\sqrt{1-\kappa^2\sin^2 q}$, for parameters
$c$ and $\kappa$ see Ref. \onlinecite{johnson-excitations-xyz}.
While  $c/J_z$ changes almost linearly between $1/2$ and $\pi/2$ as
$\alpha$ goes from  
$0$ to $1$,  $\kappa$ varies steeply from $0$ to $1$, 
such that the spinon gap $\omega_s\!=\!\omega_{min}$
becomes small already for $\alpha\!\agt\!0.5$ and 
approaches zero exponentially: $\omega_s\!\approx\! 
4c\exp\left(-\pi^2\sqrt{\alpha/8(1-\alpha)}\right)$ as
$\alpha\!\rightarrow\!1$. 
This makes the spinon spectrum ``quasi-relativistic'' 
in this regime, $\omega_q\!=\!\sqrt{(cq)^2+\omega_s^2}$. 
The spinon also becomes very light, its mass going from $m_s\!\sim\!
\alpha^{-1}$ at 
$\alpha\!\ll\!1$ to $m_s\!\sim\!\omega_s$ at $\alpha\!\agt\!0.5$, 
see Fig. \ref{fig:dispersion}(a). 
Such a transformation of the spinon spectrum strongly affects
the spinon-holon binding. Note that, when the spinon is much lighter than the holon, 
$m_s\ll\!m_h\!\simeq\!(2t)^{-1}$, the role of the latter in pairing must be 
secondary. 

The remaining question is the spinon-holon interaction.
The binding problem can be solved rigorously to order 
$O(\alpha)$. In the small-$\alpha$ regime, changes to the holon 
and the AF GSEs are of order $O(\alpha^2)$, while the spinon 
energy changes in the order $O(\alpha)$: 
$\omega_q\!=\!\omega^0\!+\!\delta\omega_q$, where
$\omega^0\!=\!J_z/2$ and  
$\delta\omega_q\!=\!\alpha J_z\cos 2q$. 
At $\alpha\!>\!0$ spinon GS momentum is $\pm\pi/2$, so the bound 
state should also have a finite momentum  $P\!=\!\pm\pi/2$, 
in agreement with the numerical data. 
Since the energy is lowered when the AF domain walls associated with 
spinon and holon cancel each other, 
the interaction can be written as a ``contact'' attraction 
of strength $V^0\!=\!-J_z/2$. This leads to a relation between interaction 
in the momentum space and spinon
dispersion, valid to order $O(\alpha)$: $V_{q,q'}\!=
\!-\omega^0\!-\!(\delta\omega_q\!+\!\delta\omega_{q'})/2$.
Using this interaction, spinon energy as above, and 
the ``bare'' holon energy $\epsilon^0_k$ in Eq. (\ref{BSE}) yields:
$\Delta\!=\!\Delta_0 (1\!-\!A\alpha)$, 
 with $\Delta_0$ given by (\ref{ising-delta}) 
and $A|\Delta_0|\!=\!J_z^2/\sqrt{16+J_z^2}$. The
slope $A$ varies from 4 to 2 for $0\!\leq\!J_z\!\leq\infty$.  
This $O(\alpha)$ result for $\Delta$ at $J_z\!=\!4.0$ is shown 
in Fig. \ref{fig:binding} by the dashed line. It is in excellent 
agreement with the numerical data in the small-$\alpha$ regime.

Based on the above analyzis, we propose the general form 
of the spinon-holon interaction 
in momentum space: $V_{q,q'}\!=\!-\sqrt{\omega_q\omega_{q'}}$. 
Using this $V_{q,q'}$, spinon energy from BA, and ``bare'' 
holon dispersion, Eq. (\ref{BSE}) is transformed into:
\begin{equation}
\label{BSE1}
1 =
-\sum_q\frac{\omega_q}{\Delta-
(\epsilon_{q+P}-\epsilon_0)-(\omega_q-\omega_0)}.  
\end{equation}
Solving this
equation gives the dependence of the binding
energy $\Delta$ on anisotropy $\alpha$ shown as a solid line
in Fig. \ref{fig:binding}.
Not only does this equation yield our small-$\alpha$ results,
but it also  
provides a very close agreement with the numerical data for all the values of 
$J_z$ and for all $\alpha$ we can access numerically. 
This agreement makes the validity of our spinon-holon
interaction ansatz very plausible.
 
As the binding energy becomes small at $\alpha\!\rightarrow\!1$, 
pairing is determined by the long-wavelength features 
of the dispersions and interaction.
Within the qualitative picture of pairing in 1D, 
both the characteristic low-energy interaction 
and the spinon mass become proportional to the 
spinon gap $\omega_s$. Thus, 
one expects $\Delta\!\sim\!-V^2m\!\sim\!-\omega_s^3\!\sim\!-\exp\left(
-3\pi^2/\sqrt{8(1-\alpha)}\right)$. 
We can derive this behavior from Eq. (\ref{BSE1}) explicitly: 
$\Delta\!\approx\!-{\cal D}(J_z,\alpha)\omega_s^3$, 
where the exponential behavior 
is determined solely by the spinon gap. 
The holon energy scale is secondary as it only enters 
the ``regular'' pre-factor ${\cal D}(J_z,\alpha)$.
Altogether, this explains the quick (exponential) drop-off of $\Delta$ at  intermediate $\alpha$.

One can see from the asymptotic expression that the binding energy
vanishes in the isotropic limit together with the spinon gap. Thus,
our spinon-holon interaction ansatz provides a natural and simple
explanation of the non-zero binding but no bound state at
$\alpha\!=\!1$: it is possible because the interaction of the holon 
with the long-wavelength spinon vanishes together with the spinon
energy. Then the pairing is not strong enough to produce a
bound state.
 
\emph{Conclusions.}\
To summarize, we have studied the spinon-holon interaction in the
anisotropic $t$-$J$ model. We have demonstrated that the finite-size
extrapolation of the ED and DMRG data of very high accuracy is
possible and it can provide reliable values of the spin-holon binding
energy. These data are in excellent agreement with the theory based on
the BSE with the contact spinon-holon interaction. The theory provides
a coherent, simple, and consistent explanation of the binding 
for all values of anisotropy. Within the offered picture, the holon is
attracted to the ``slow'' spinon in the $\alpha\!\rightarrow\!0$ limit
while at $\alpha\!\agt\!0.5$ the ``fast'' spinon can only be bound weakly.
We also offer an explanation of the non-zero attraction but no bound
state in the isotropic $t$-$J$ model.

\emph{Acknowledgments.} 
We would like to thanks to A. Bernevig and  O. Starykh for 
fruitful discussions.
This work was supported in part by DOE grant
DE-FG02-04ER46174, by the Research Corporation  
(J.\v{S}. and A.L.C.), and by NSF grant DMR-0605444 (S.R.W.).


\begin{thebibliography}{99}

\bibitem{LiebWu} 
E. H. Lieb and F. Y. Wu, Phys. Rev. Lett. {\bf 20}, 1445 (1968).

\bibitem{Bernevig}
B. A. Bernevig \emph{et al.}, Phys. Rev. B {\bf 65}, 195112 (2002).

\bibitem{sorella-1d} 
S. Sorella and A. Parola, Phys. Rev. B {\bf 57}, 6444 (1998).

\bibitem{dmrg} 
S. R. White, Phys. Rev. Lett. {\bf 69}, 2863 (1992);
S. R. White, Phys. Rev. B {\bf 48}, 10345 (1993).

\bibitem{singlesite} 
S. R. White, Phys. Rev. B {\bf 72}, 180403 (2005).

\bibitem{shiba-ogata-review} 
H. Shiba and M. Ogata, Prog. Theor. Phys. Supp. {\bf 108}, 265 (1992).

\bibitem{zotos-tbc} 
X. Zotos \emph{et al.}, Phys. Rev. B {\bf 42}, 8445 (1990).

\bibitem{yang-xxz} 
C. N. Yang and C. P. Yang, Phys. Rev. {\bf 150}, 321 (1966).

\bibitem{johnson-excitations-xyz}
J. D. Johnson \emph{et al.}, Phys. Rev. A {\bf 8}, 2526 (1973).
                                                                                
\bibitem{elsewhere}
Further technical details will be presented elsewhere.
                                                                              
\bibitem{vega-woynar-size-corr} 
H. J. de Vega and F. Woynarovich, Nucl. Phys. B {\bf 251}, 439 (1985).

\bibitem{LL} 
V. B. Berestetskii \emph{et al.}, \emph{Quantum Electrodynamics},
vol. 4 of \emph{Landau and Lifshitz Course of Theoretical Physics} 
(Pergamon, New York, 1970), pp. 552 -- 559.

\end{thebibliography}
\end{document}